\documentclass[twocolumn%
              ,superscriptaddress%
              ,aps%
              ,prd%
              ,longbibliography
              ]{revtex4-1}

\usepackage{amsmath,amssymb}
\usepackage{graphicx}

\begin{document}

\title{Semiclassical evolution in phase space for a softly chaotic system}


\author{Gabriel M. Lando}
\affiliation{Centro Brasileiro de Pesquisas F\'isicas,
Rua Xavier Sigaud 150, 22290-180, Rio de Janeiro, R.J., Brazil}
\author{Alfredo M. Ozorio de Almeida}
\affiliation{Centro Brasileiro de Pesquisas F\'isicas,
Rua Xavier Sigaud 150, 22290-180, Rio de Janeiro, R.J., Brazil}
\date{\today}

\begin{abstract}
An initial coherent state is propagated exactly by a kicked quantum Hamiltonian and its associated classical stroboscopic map. The classical trajectories within the initial state are regular for low kicking strengths, then bifurcate and become mainly chaotic as the kicking parameter is increased. Time-evolution is tracked using classical, quantum and semiclassical Wigner functions, obtained via the Herman-Kluk propagator. Quantitative comparisons are also included and carried out from probability marginals and autocorrelation functions. Sub-Planckian classical structure such as small stability islands and thin/folded classical filaments do impact semiclassical accuracy, but the approximation is seen to be accurate for multiple Ehrenfest times.
\end{abstract}

\maketitle  

\section{Introduction}

The extreme sensitivity to initial conditions, a fundamental feature of chaotic dynamics, can be prevalent over all of phase space or be limited to bounded, specific regions, which are shared with regular orbits. The former case, known as \emph{hard chaos} \cite{Gutzwiller1990}, has been extensively studied and connected to quantum mechanics through random matrix theory (RMT) \cite{Bohigas1984, Haake2018, Ozorio1990}; while the latter, vaguely dubbed as \emph{soft chaos}, presents additional difficulties: Here, full agreement with RMT is unfeasible due to the persistence of regular orbits, while at the same time it is hard to implement periodic orbit theory due to structure being too heterogeneous.

Nevertheless, the semiclassical features of softly chaotic systems was investigated in several papers. Employing an adapted Herman-Kluk propagator \cite{Heller1981, Herman1984}, Schoendorff \emph{et al} reproduced the quasi-energy spectrum of a periodically driven quantum system with remarkable accuracy \cite{Schoendorff1998}, while Maitra used a similar technique to evaluate the propagation of momentum eigenstates for the quantized standard map \cite{Chirikov1979} in various kicking regimes \cite{Maitra2000}. The accuracy of semiclassical approximations for regular and chaotic dynamics was also investigated by Kaplan \cite{Kaplan2004}, showing that regular orbits might actually be more troublesome than chaotic ones. In the time domain, the situation is the same as mentioned earlier: Semiclassical approximations seem to work well for both short and long times, but not always for the intermediate regime. 

A comprehensive study of what effectively happens in phase space is still missing, however. The fundamental characteristic of soft chaos, namely the existence of stability islets, has been associated to failures in semiclassical propagation when their areas were smaller than $\hbar$ \cite{Maitra2000}, but this was hardly ever explored outside autocorrelation functions. A phase-space description of semiclassical propagation should certainly facilitate classical-quantum comparisons and lead to insights regarding, for instance: a) what are the islet sizes and where are they located; b) does the attraction/scattering of the initial distribution by elliptic/hyperbolic fixed points impact semiclassics; c) what is the time scale for which the classical orbits are able to faithfully reproduce quantum behavior; d) does sub-Planckian structure really affect the validity of the approximations?

We here attempt to address such points. Following Berry \emph{et al} \cite{Berry1979}, we devise a stroboscopic map \emph{on the plane}, not on the torus or cylinder as Chirikov's \cite{Chirikov1979}, for which chaotic orbits are either bounded by regular regions or diverge very slowly -- we call it the ``coserf map". As in some previous investigations \cite{Schoendorff1998,Maitra2000}, we use the Herman-Kluk semiclassical propagator to approximate the exact quantum time-evolution of a coherent state. As an initial value representation \cite{Brumer1994, Miller2001}, the Herman-Kluk propagator is not affected by caustics and does not requite the trajectory selection of standard semiclassical methods \cite{Heller1991}. This, together with the boundedness of the coserf map's chaotic trajectories, eliminates any need to discern between regularity and chaos and allows us to carelessly integrate over arbitrary orbits. 

Special attention is payed to defining a meaningful time scale: The Ehrenfest time, with which we quantify the notion of ``short" and ``long" time regimes, is extracted as the instant at which the quantum and classical autocorrelation functions start to disagree. We then track and visually compare the classical, semiclassical and exact quantum evolutions of the Wigner function of a coherent state with respect to this time scale, using marginals and autocorrelation fuctions for quantitative comparisons. The trajectories used to propagate the initial state present all possible topologies as a function of the map's kicking strength, going smoothly from regularity to chaos. Its classical evolution is, therefore, a mixture of the regions captured by regular orbits and the distorted portions subject to the intertwined chaotic dynamics. Structure with areas smaller then $\hbar$ are ubiquitous and easily identifiable for the majority of the chosen time regimes.

The text is organized as follows: Secs.~II and III are reviews of discrete maps on the plane and their quantum equivalents, respectively. In Sec.~IV we adapt the HK propagator to discrete times. We proceed to Sec.~V, where we perform the numerical analysis of the coserf map, after which a discussion follows in Sec.~VI. The final remarks are then presented in Sec.~VII, which is followed by two appendices.  

\section{Classical Maps}

Due to complete integrability, continuous time-independent Hamiltonian systems cannot be chaotic for one degree of freedom \cite{Gutzwiller1990}. For more freedoms chaotic behavior starts to become the rule rather than the exception, but an at least 4-dimensional phase space is now required -- not helping at all with visualization. However, for time-dependent Hamiltonians chaos can be reached within a 2-dimensional phase space. Kicked systems display a particular type of time-dependence that renders their dynamics discrete, in connection with Poincar\'e sections of larger continuous systems \cite{Gutzwiller1990,Ozorio1990,Berry1979}. The corresponding stroboscopic map is our main classical tool and all quantities associated to it can be written exactly, as we shall develop in the following subsections.

\subsection{Discrete dynamics}

We begin by defining the one-dimensional, time-dependent Hamiltonian 
\begin{align}
H(q,p;t) = \frac{p^2}{2} + T V(q) \sum_k \delta (t - T k) \, , \quad n \in \mathbb{N} \, ,\label{eq:timeham}
\end{align}
where $q$ is the position, $p$ is the momentum, $V(q)$ is the position-dependent potential energy and $t$ is time. The sum of delta functions expresses the fact that the potential energy is turned on at times multiple of the ``kicking strength" $T$, \emph{i.e.} we can define a discretized time
\begin{equation}
\tau_k = T k \, , \quad k \in \mathbb{N} \, , \label{eq:disctime}
\end{equation}
outside of which the system evolves freely with constant momentum $p$. Effectively, \eqref{eq:timeham} describes a particle subject to purely kinetic and potential kicks \cite{Berry1979, Maitra2000, Tabor1983}, for which Hamilton's equations of motion reduce to the finite difference map 
\begin{align}
\begin{cases}
q_{i} = q_{i-1} + T p_{i-1} \\
p_{i} = p_{i-1} - T V'(q_{i}) 
\end{cases} , \quad  i \in \mathbb{N}, \label{eq:sys}
\end{align}  
with time average 
\begin{align}
\bar{H}(q,p) = \frac{p^2}{2} + V(q) \, . \label{eq:ham}
\end{align}
The phase-space point $(q_k, p_k)$ is obtained by subjecting $(q_0,p_0)$ to $k$ iterations of the map \eqref{eq:sys}, which can also be seen as a finite difference version of the composition of two distinct Hamiltonian half--evolutions, generated by $H_{1/2}(p)$ and $H_1(q)$:
\begin{align}
\begin{cases}
H_\frac{1}{2}(p) = \dfrac{p^2}{2} &\Rightarrow \quad \!\!
\begin{cases}
q_{i-\frac{1}{2}} = q_{i-1} + T p_{i-1} \\
p_{i-\frac{1}{2}} = p_{i-1}
\end{cases}  \\[12pt]
H_1(q) = \, V(q) &\Rightarrow \quad \!\!
\begin{cases}
q_{i} = q_{i-\frac{1}{2}}  \\
p_{i} = p_{i-\frac{1}{2}} - T V'( q_{i-\frac{1}{2}} )
\end{cases} 
\end{cases} \!\!\!\!\!, \label{eq:split}
\end{align}
easily seen to be equivalent to \eqref{eq:sys} by direct substitution. 

The equivalent maps \eqref{eq:sys} and \eqref{eq:split} are, in practice, used in the limit of small $T$ to approximate solutions of continuous systems, being the most elementary representatives of a class of numerical methods known as splitting symplectic integrators (SSIs). For the purpose of numerical integration, more elaborate splitting algorithms have been devised which give birth to extremely accurate and efficient SSIs \cite{Yoshida1990, Atela1992}, especially for high dimensional systems \cite{Tselios2013, Blanes2013}. Our focus here, however, is to consider such a discretization as a system on its own. 

\subsection{Discretized lagrangian and corresponding action}

We define a discrete Lagrangian function for the iteraction from $i-1$ to $i$ as
\begin{align}
L_{i}(q_{i}, q_{i-1}) &= \frac{1}{2} \left( \frac{q_{i} - q_{i-1}}{T} \right)^2 - V(q_{i}) \, ,
\end{align}
with correspondent discrete action
\begin{align}
S_{i} (q_{i}, q_{i-1}) &= \frac{(q_{i} - q_{i-1})^2}{2 T} - T V(q_i) \, . \label{eq:action} \\
&= T L_{i}(q_{i}, q_{i-1}) \, .
\end{align}
It is easy to see that $S_{i}$ fulfills everything that is expected from a discretization of the continuous action. Namely, just as a continuous generating function, 
\begin{align}
p_{i} = \frac{\partial S_{i}}{\partial q_{i}} \quad \text{and} \quad p_{i-1} = - \frac{\partial S_{i}}{\partial q_{i-1}} \, ;
\end{align}
and, by defining the total action from $(q_0, p_0)$ to $(q_{k}, p_{k})$ as 
\begin{align}
S(q_0, q_{k}) &= \sum_{i=0}^k S_{i} (q_{i}, q_{i-1}) \, , \label{eq:discaction}
\end{align}
we have the required Riemann sum correspondence 
\begin{align}
S(q_0, q_{k}) &= \sum_{i=0}^k T L_{i}(q_{i}, q_{i-1}) \\
&\stackrel{T \to 0}{\approx} \int_0^t d \tau L_\tau (q_\tau, \dot{q}_\tau) \, ,
\end{align}
where $L_\tau (q_\tau, \dot{q}_\tau)$ is the continuous Lagrangian function.

We point out that our action in \eqref{eq:action} is not the same as the one introduced in \cite{Tabor1983}. The reason for discrepancy is that the map 
\begin{align}
\begin{cases}
p_{i} &= p_{i-1} - T V'(q_{i-1}) \\
q_{i} &= q_{i-1} + T p_{i} 
\end{cases} \, , \quad  i \in \mathbb{N}, \label{eq:systabor}
\end{align}  
used in \cite{Tabor1983}, is \emph{not} the same as our map in \eqref{eq:sys}. As $T \to 0$ the difference between different ordering choices also goes to zero, since both maps reproduce the level curves of \eqref{eq:ham}. For finite $T$, however, noticeable differences are present. This will be very important when we discuss the quantization of \eqref{eq:sys} and Wigner dynamics in Sec. III. 

\subsection{The monodromy matrix}

The monodromy (or stability \cite{Garashchuk2000, Ushiyama2001}) matrix, given in the one dimensional case by 
\begin{align}
M_t(q_0, p_0) = 
\begin{pmatrix}
\dfrac{\partial q_t (q_0,p_0)}{\partial q_0} & \dfrac{\partial q_t(q_0,p_0)}{\partial p_0} \\[10pt]
\dfrac{\partial p_t (q_0,p_0)}{\partial q_0} & \dfrac{\partial p_t(q_0,p_0)}{\partial p_0}   
\end{pmatrix} \, ,
\end{align}
is a fundamental object in semiclassical approximations, since virtually all semiclassical propagators have amplitudes expressed as functions of its components. To obtain it, the usual procedure involves solving the differential equation  
\begin{align}
\dot{M}_t = J H'' M_t \, , \quad M \big|_{t=0} = M_0 \, , \label{eq:mon}
\end{align}
where $H''$ is the Hamiltonian's Hessian matrix and
\begin{align}
J = 
\begin{pmatrix}
0 & 1 \\
-1 & 0
\end{pmatrix} \, .
\end{align}
Since \eqref{eq:mon} is linear, it has a single solution and can itself be taken as defining the monodromy matrix, being usually solved by finite differences or more elaborate methods \cite{Garashchuk2000, Ushiyama2001, Swenson2011}. As we shall see, for the case of discrete maps the monodromy matrix is exact, being expressible as a very simple product \cite{Tabor1983, Maitra2000}.

We start by noticing that the Jacobians for each step in \eqref{eq:split} are the symplectic matrices
\begin{align}
\frac{\partial (q_{i-\frac{1}{2}} , p_{i-\frac{1}{2}})}{\partial (q_{i-1}, p_{i-1})} &= 
\begin{pmatrix}
1 & T \\
0 & 1
\end{pmatrix} \label{eq:jac1} \\
\frac{\partial (q_{i}, p_{i})}{\partial (q_{i-\frac{1}{2}} , p_{i-\frac{1}{2}})} &= 
\begin{pmatrix}
1 & 0 \\
-T V''(q_{i-\frac{1}{2}}) & 1
\end{pmatrix} \, , \label{eq:jac2} 
\end{align}
such that the Jacobian matrix for the composite iteration \eqref{eq:split}, by the chain rule, is the same as for \eqref{eq:sys}:
\begin{align}
\frac{\partial (q_{i}, p_{i})}{\partial (q_{i-\frac{1}{2}} , p_{i-\frac{1}{2}})} \frac{\partial (q_{i-\frac{1}{2}} , p_{i-\frac{1}{2}})}{\partial (q_{i-1}, p_{i-1})}  &= 
\frac{\partial (q_{i}, p_{i})}{\partial (q_{i-1} , p_{i-1})}  \, . \label{eq:chain}
\end{align}
Therefore, we can express the monodromy associated to $k$ kicks as a function of the initial values $(p_0,q_0)$ as
\begin{align}
M_{k} (p_0,q_0) &= \frac{\partial (p_{k}, q_{k})}{\partial (p_0, q_0)} \notag \\
&= \frac{\partial (p_1, q_1)}{\partial (p_0, q_0)} \frac{\partial (p_2, q_2)}{\partial (p_1, q_1)} \dots \frac{\partial (p_{k}, q_{k})}{\partial (p_{k-1}, q_{k-1})} \notag \\ 
\Rightarrow M_{k} (p_0,q_0)&= \prod_{i=1}^k \frac{\partial (p_{i}, q_{i})}{\partial (p_{i-1}, q_{i-1})} \, , \quad M_0 = I \, , \label{eq:discmon}
\end{align}
which, by \eqref{eq:chain}, is the product of the Jacobians in \eqref{eq:jac1} and \eqref{eq:jac2} from 0 to $k$. Since the product of symplectic matrices is symplectic, it is clear that the final monodromy is symplectic. Such a decomposition has the further advantage that it can be computed in parallel with the equations of motion, decreasing computational cost. 

\section{Quantum maps}

Classical systems that possess exact quantizations, such as in \cite{Lando2019}, are perfect to test semiclassical methods. However, since dealing with chaotic trajectories is so much easier when they do not diverge, most of quantum chaology has been focused on compact or partially compact phase spaces (such as the torus or the cylinder). Such systems present the sometimes undesirable characteristic that evolution, when displayed on the plane, looks discontinuous: We end up with sliced states that, although glued via equivalence relations at the boundaries, are non-intuitive to the eye -- especially if one considers interference patterns. 

The maps generated from Hamiltonians of the form \eqref{eq:timeham} have an exact quantum analogue. As expected of maps on the plane, they present no accessible hard chaotic limit: trajectories will simply diverge faster as $T$ is increased and are not bounded to any phase-space region. However, since we are not interested in the hard regime, analyzing soft chaos in the open plane allows us to track the initial state as it continuously folds and deforms, providing a beautifully intuitive picture of classical propagation and evoking comparisons with its quantum counterpart. 

\subsection{Quantum kicked systems}

Since the Hamiltonian in \eqref{eq:ham} is time-independent, its corresponding time-evolution operator is given by
\begin{align}
\exp \left[ -\left(\frac{i t}{\hbar} \right) \hat{H} \right] = \exp \left[ - \left( \frac{i t}{\hbar} \right) \left( \frac{\hat{p}^2}{2} + \hat{V}(\hat{q}) \right) \right] \, . \label{eq:exact}
\end{align}
Due to the non-commutativity of $\hat{p}$ and $\hat{q}$, the splitting
\begin{align}
\hat{U} = \exp \left[ - \left( \frac{i t}{\hbar} \right) \hat{V}(\hat{q}) \right] \exp \left[ - \left( \frac{i t}{\hbar} \right) \frac{\hat{p}^2}{2} \right] \label{eq:qkick}
\end{align}
is not equivalent to the RHS of \eqref{eq:exact}, with energy errors of $\mathcal{O}(t^1)$ \cite{Yoshida1990}. The RHS of \eqref{eq:qkick} is, however, the \emph{exact} quantization of the time-dependent Hamiltonian in \eqref{eq:timeham}, as long as the time $t$ is identified with the kicking period $T$ \cite{Berry1979}. This identification is clearly consistent with viewing $T$ as a kicking strength used to iterate the system from $i-1$ to $i$, the final state being obtained by $k$ iterations of \eqref{eq:qkick} on the initial state. In analogy to \eqref{eq:sys}, the system's wavefunction at $i$ can then be written recursively as
\begin{equation}
\langle q \vert \psi_{i} \rangle = \int dq' \langle q \vert \hat{U}_1 \vert q' \rangle \langle q \vert \psi_{i-1} \rangle \, , \quad i \in \mathbb{N} \, ,\label{eq:qkick2}
\end{equation}
where $\hat{U}_1$ represents the operator for a single kick. The position representation of $\hat{U}_1$ in the integral above has the very simple form \cite{Berry1979}
\begin{align}
\langle q \vert \hat{U}_1 \vert q' \rangle &= \left( \frac{1}{2 \pi \hbar T} \right)^\frac{1}{2} \exp \left( - \frac{i \pi}{4} \right)  \notag \\
&\qquad \times \exp \left[ -\frac{i T V(q)}{\hbar}   + \frac{i (q-q')^2}{2 \hbar T }\right] \, . \label{eq:qpos}
\end{align}
Thus, the full propagator for $k$ steps is given by
\begin{align}
	\langle q \vert \hat{U}_k \vert q' \rangle &= \int dq_{k-1} \, dq_{k-2} \dots dq_1 \langle q \vert \hat{U}_1 \vert q_{k-1} \rangle \notag \\
	&\qquad \langle q_{k-2} \vert \hat{U}_1 \vert q_{k-1} \rangle \dots \langle q_1 \vert \hat{U}_1 \vert q' \rangle \, , \label{eq:propk}
\end{align}
and, for the final wavefunction,
\begin{align}
\langle q \vert \psi_k \rangle &= \int dq' \langle q \vert \hat{U}_k \vert q' \rangle \langle q' \vert \psi_0 \rangle \, . \label{eq:qk}
\end{align}

The ordering in \eqref{eq:qkick}, with the momentum operator up front, is chosen in parallel with the classical maps \eqref{eq:sys} and \eqref{eq:split}, where the position translation is performed first. This is the correct correspondence, since the momentum operator is the generator of translations in position representation. If we had chosen to translate first in momentum as in the classical map \eqref{eq:systabor}, in direct correspondence with reversing the ordering of operators in \eqref{eq:qkick}, straightforward calculations show that \eqref{eq:qpos} would have to be slightly modified, $V(q)$ now becoming $V(q')$. Although small, this modification in the quantum propagator is responsible for creating significant differences in the final wavefunctions, just as the map \eqref{eq:sys} is significantly different from \eqref{eq:systabor}.

\subsection{The Wigner function and time discretization}

Wavefunctions and probability marginals provide fair comparisons between semiclassical and quantum calculations, but are virtually useless in establishing quantum-classical analogies. As usual, to gain intuition about a quantum system's classical backbone it is necessary to move to phase space. The Wigner function 
\begin{align}
W_t(q,p) =  \frac{1}{\pi \hbar} \! \int \! d\tilde q\langle q+\tilde q | \psi_t \rangle\langle\psi_t | q-\tilde q\rangle
e^{-2i\tilde q p/ \hbar}\,, \label{eq:wig}
\end{align}
is, then, the fundamental object: Both position and momentum marginals can be easily extracted from it by integration, \emph{e.g.},
\begin{equation}
\vert \langle q \vert \psi_t \rangle \vert^2 = \int dp \, W_t (q,p) \, , \label{eq:marginal}
\end{equation}
as well as the autocorrelation function, given by
\begin{align}
C_t =  \sqrt{2 \pi \int dp \, dq \,W_t(q,p) W_0 (q,p)} \, \, . \label{eq:autoclass}
\end{align}
The Moyal formulation of quantum mechanics presents yet another interesting Wigner function property: Its time-evolution can be written as a series in $\hbar$, the first term being Lioville evolution:
\begin{align}
\frac{\partial W_t (q,p)}{\partial t} = - \left\{ W_t (q,p), H(q,p) \right\} + \mathcal{O}(\hbar) \, , \label{eq:wigclass}
\end{align}
where $H(q,p)$ is the classical Hamiltonian and the curly brackets are Poisson's. Two aspects brought up by the equation above deserve some elaboration: Firstly, since the Wigner function's time-evolution can be seen as a series in $\hbar$, we can semiclassically interpret quantum fringes as terms of order higher or equal to $\hbar$ arising from the classical Wigner function interfering with itself \cite{Ozo1998,Lando2019}; Secondly, we see that the Wigner function evolves as a probability distribution, \emph{i.e.}~it is obtained from propagating trajectories backwards in time due to the minus sign in \eqref{eq:wigclass}. For the discrete maps introduced in the previous section, however, we notice that backward propagation reverses the order of shears: Shearing first in $q$ is replaced by shearing first in $p$. In summary, when studying the map \eqref{eq:sys}, we must propagate the Wigner function using its twin map \eqref{eq:systabor}.

As of time discretization, since the integration in \eqref{eq:wig} is performed with respect to a continuous variable, adapting it (and all of the Wigner function properties) from the $T \to 0$ to the finite $T$ case is done by simply substituting $t \mapsto k$. For $k=0$ we use the system's initial state, while for $k > 0$ we employ \eqref{eq:sys} and \eqref{eq:qkick2} for the classical and quantum regimes, respectively.

\section{Initial Value Representations}

In their original form \cite{Littlejohn1992}, semiclassical formulas required selecting trajectories based on boundary conditions -- a procedure that may not be feasible, especially when dealing with multiple degrees of freedom. An even worse aspect is that the standard propagation diverges along \emph{caustics}, which are generalized return points that become unavoidable as time grows \cite{Lando2019, Littlejohn1992}. Although the seminal work from Tomsovic and Heller showed that the van Vleck-Gutzwiller propagator was able to reproduce quantum behavior from classically chaotic orbits for the stadium billiard \cite{Heller1991}, the root-search and caustic avoidance needed are the most computationally expensive steps in the calculations, and some trajectories end up being discarded. Initial/Final Value Representations (IVR/FVRs), on the other hand, replace sums over trajectories by an integral over all initial positions and momenta, avoiding root-search problems \cite{Kay1994(2)}. As an off-shot, IVR/FVR amplitudes tend to zero at caustics, allowing integration over problematic trajectories without any special procedure. 

\subsection{The Herman-Kluk propagator}

Using a time-dependent, quadratic Hamiltonian to propagate static Gaussian packets, the Frozen Gaussian Approximation (FGA) \cite{Heller1981} has become the method of choice for chemists. The Herman-Kluk (HK) propagator \cite{Herman1984}, particularly, is unarguably the most employed method in the field, although many theoretical aspects about it remain obscure and have created a great deal of confusion in the last decade (see \cite{Baranger2001} and subsequent comments, for example). As a typical IVR, it is represented by an integral over initial position and momentum,
\begin{align}
\langle q \vert \hat{U} \vert q' \rangle_{\text{HK}} &= \frac{1}{2 \pi \hbar} \int dq_0 \, dp_0 \, R_t(q_t,p_t) \langle q' \vert q_t p_t \rangle \notag \\
&\qquad \qquad \langle q_0 p_0  \vert q \rangle \exp \left[ \frac{i}{\hbar} S_t (q_0,p_0) \right] \, , \label{eq:HK}
\end{align}
where we use Herman and Kluk's original definitions for a system with one degree of freedom:
\begin{align}
R_t (q_t,p_t) &=  \sqrt{\frac{1}{2} \left[ \frac{\partial p_t}{\partial p_0} + \frac{\partial q_t}{\partial q_0} + \frac{i}{\hbar} \frac{\partial p_t}{\partial q_0} - i \hbar \frac{\partial q_t}{\partial p_0} \right]} \\[6pt]
\langle q \vert q_t p_t \rangle \! &= \! \left( \frac{1}{\pi} \right)^\frac{1}{4} \!\!\! \exp \left\{ - \frac{1}{2} ( q - q_t)^2 + \!\! \frac{i p_t }{\hbar} (q - q_t) \right\} \\[6pt]
S_t(q_0,p_0) &  = \int_0^t d\tau \left[ p_\tau \frac{\partial q_\tau}{\partial \tau} - H(q_\tau,p_\tau) \right] \, ,
\end{align}
where $H(q_t, p_t)$ is the classical Hamiltonian and $(q_t, p_t)$ are solutions to
\begin{align}
\begin{cases}
\dfrac{d q}{d t} = \dfrac{\partial H}{\partial p} \\[8pt]
\dfrac{d p}{d t} = - \dfrac{\partial H}{\partial q}
\end{cases} , \quad  (q, p) \big|_{t = 0} = (q_0, p_0).
\end{align} 
Since the monodromy matrix is  never singular, it is trivial to see that the $R_t$ pre-factor in the HK propagator is never zero. This does \emph{not} mean, as is sometimes stated in literature, that the propagator is free from caustics: All IVRs are ``free from caustics" in the sense that they do not diverge along them, but acquired phases due to the change of metaplectic sheets are still present \cite{Ozorio2014, Littlejohn1992}. For the HK propagator these sheets are poorly understood, but acquired phases cannot be simply dismissed. Here, the Maslov indexes associated to changing sheets appear as a need to keep track of the correct branch of the complex square root in the pre-factor. As has been pointed out by Kay \cite{Kay1994}, this procedure amounts exactly to the usual caustic counting present in other semiclassical propagators \cite{Littlejohn1992}.

\subsection{The Herman-Kluk propagator for discrete times}

The semiclassical wavefunction propagated by the HK method is just \eqref{eq:qk} with the approximation \eqref{eq:HK} for $\langle q \vert \hat{U}_k \vert q' \rangle$, instead of the exact expression \eqref{eq:propk}. The passage from the continuous-time to the discrete-time version is done by simply substituting the action \eqref{eq:discaction} and the monodromy \eqref{eq:discmon} in its defining formula \eqref{eq:HK}. We choose the correct square root branch by employing a technique described by Swenson \cite{Swenson2011}, in which we evaluate the HK pre-factor at each kick $i$ and the Maslov index associated to an orbit increases by one if the conditions
\begin{align}
\begin{cases}
\qquad \qquad \qquad \Re \left[ R_{i}(q_{i}, p_{i}) \right]  &< 0   \\
\Im(\left[ R_{i}(q_{i}, p_{i}) \right] \times \Im \left[ R_{i-1}(q_{i-1}, p_{i-1}) \right] &< 0
\end{cases} 
\end{align}
are simultaneously fulfilled. 

We emphasize that we do not deal with a discretization of the HK propagator: Positions and momenta are still taken as continuous (although, numerically, they are obviously a mesh). The discretized quantity is time, taken as $k$ kicks with kicking strength $T$. By taking $T \ll 1$ we can recover the continuous time limit (and, consequently, erase chaos from the system).

\section{Numerical Testing}

The numerical analysis of chaotic systems is often marred by trajectories that escape to infinity \cite{Berry1979,Tomsovic1991,Brumer2000}. Even though IVRs are uniform approximations and do not mind caustics, they still blow up when divergent chaotic trajectories are present. In order to avoid this technical difficulty, semiclassical dynamics usually focuses on maps on the torus, billiards or systems that present very slow divergence speeds \cite{Berry1979, Heller1991}. For IVRs such as Heller's thawed Gaussian approximation (TGA), divergent trajectories in a chaotic system are not a problem, since the TGA's amplitude goes to zero as the separation between classical trajectories tends to infinity \cite{Liberto2016}. However, it is somewhat clear that Heller's TGA and some of its relatives are inferior to Herman and Kluk's \cite{Kay1994}, even though the latter includes a possibly divergent amplitude. In order to avoid infinities, we follow \cite{Berry1979} and choose a Hamiltonian that, when discretized in the form \eqref{eq:sys}, presents chaotic orbits that are bounded between regular phase-space regions.

Our semiclassical analysis will be focused on the propagation of minimal-uncertainty wavepackets, \emph{i.e.} coherent states, labeled by a complex parameter $\alpha$. The wavefunction for a coherent state centered at $(q,p) = (\Re(\alpha), \Im(\alpha))$, where $\Re (\alpha)$ and $\Im (\alpha)$ are the real and imaginary parts of $\alpha$, will be taken as
\begin{align}
\langle q \vert \alpha \rangle \!=\! \left( \frac{1}{\pi} \right)^\frac{1}{4} \!\!\! \exp \left\{ - \frac{1}{2} \left[ q - \Re(\alpha)  \right]^2 \!\! + i \Im(\alpha) \left[ q - \Re(\alpha) \right] \right\} .  \label{eq:coh} 
\end{align}
In the definition above we have adopted the convention $\hbar=1$, which shall be used throughout this section.

\subsection{The classical coserf map}

The time-dependent version \eqref{eq:timeham} of the averaged ``coserf" Hamiltonian
\begin{align}
\bar{H}_{\text{coserf}}(q,p) &= \frac{p^2}{2} + V_{\text{coserf}}(q) \, , \label{eq:erfcosham}
\end{align}
with
\begin{align}
V_{\text{coserf}}(q) &=  \frac{q^2}{2} - 2 \cos(q) - \frac{\sqrt{\pi} \, \text{erf}(q)}{2}   \, , \label{eq:erfcosv}
\end{align}
generates the discrete map
\begin{align}
\begin{cases}
q_{i} = q_{i-1} + T p_{i-1} \\
p_{i} = p_{i-1} - T \left[ q_i + 2 \sin q_i - \exp (-q_i^2) \right] 
\end{cases} \!\!\! \, .\label{eq:erfcosmap}
\end{align} 
Its reverse ordering twin is given by
\begin{align}
\begin{cases}
p_{i} = p_{i-1} - T \left[ q_{i-1} + 2 \sin q_{i-1} - \exp (-q_{i-1}^2) \right]  \\
q_{i} = q_{i-1} + T p_i 
\end{cases} \!\!\!\!\!\!\! \, . \label{eq:erfcosmap2}
\end{align} 
The orbits of \eqref{eq:erfcosmap} and \eqref{eq:erfcosmap2} are visually identical to the level curves of \eqref{eq:erfcosham} for $T \ll 1$ . However, we expect drastic changes as $T$ is increased. In Fig.~1 we introduce the coserf map by choosing some representative trajectories lying within the Wigner function of an initial coherent state, following their orbits as a function of $T$. 

\begin{figure}
	\includegraphics[width=\linewidth]{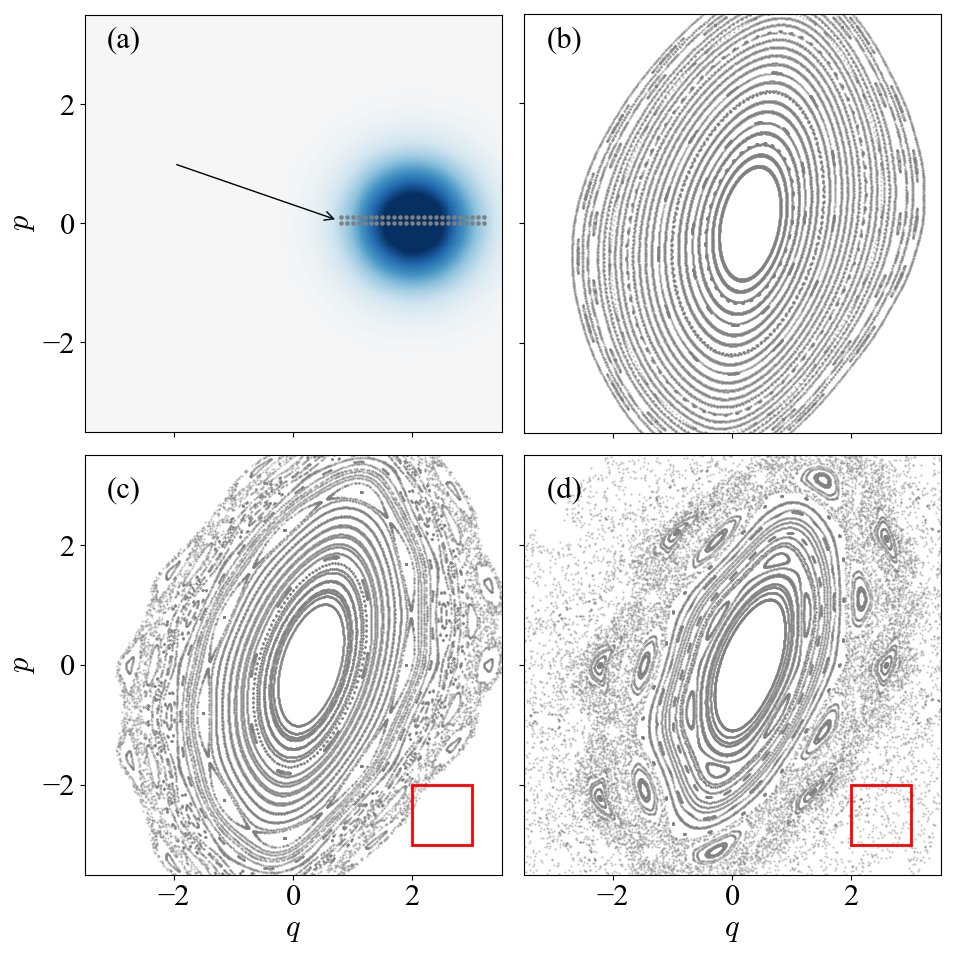}
	\caption{\textbf{(a)} We choose a small set of initial points forming a narrow strip (see arrow) and lying within the Wigner function (blue) of an initial coherent state centered at $(q_0=2, p_0=0)$. Each point in the strip forms an orbit as it is evolved according to \eqref{eq:erfcosmap2}, which will be later used to classically propagate the full Wigner evolution. Three kicking strengths are chosen: In \textbf{(b)} we have $T=0.3$, for which the orbits are distorted but remain mostly regular; In \textbf{(c)} we choose $T=0.447$, for which the map displays regular, bifurcated and weakly chaotic orbits; For $T=0.581$, shown in \textbf{(d)}, the majority of orbits has become chaotic. The red squares in \textbf{(c)} and \textbf{(d)} have an area equal to $\hbar=1$.}
	\label{fig:alpha}
\end{figure}

\begin{figure*}
	\includegraphics[width=\linewidth]{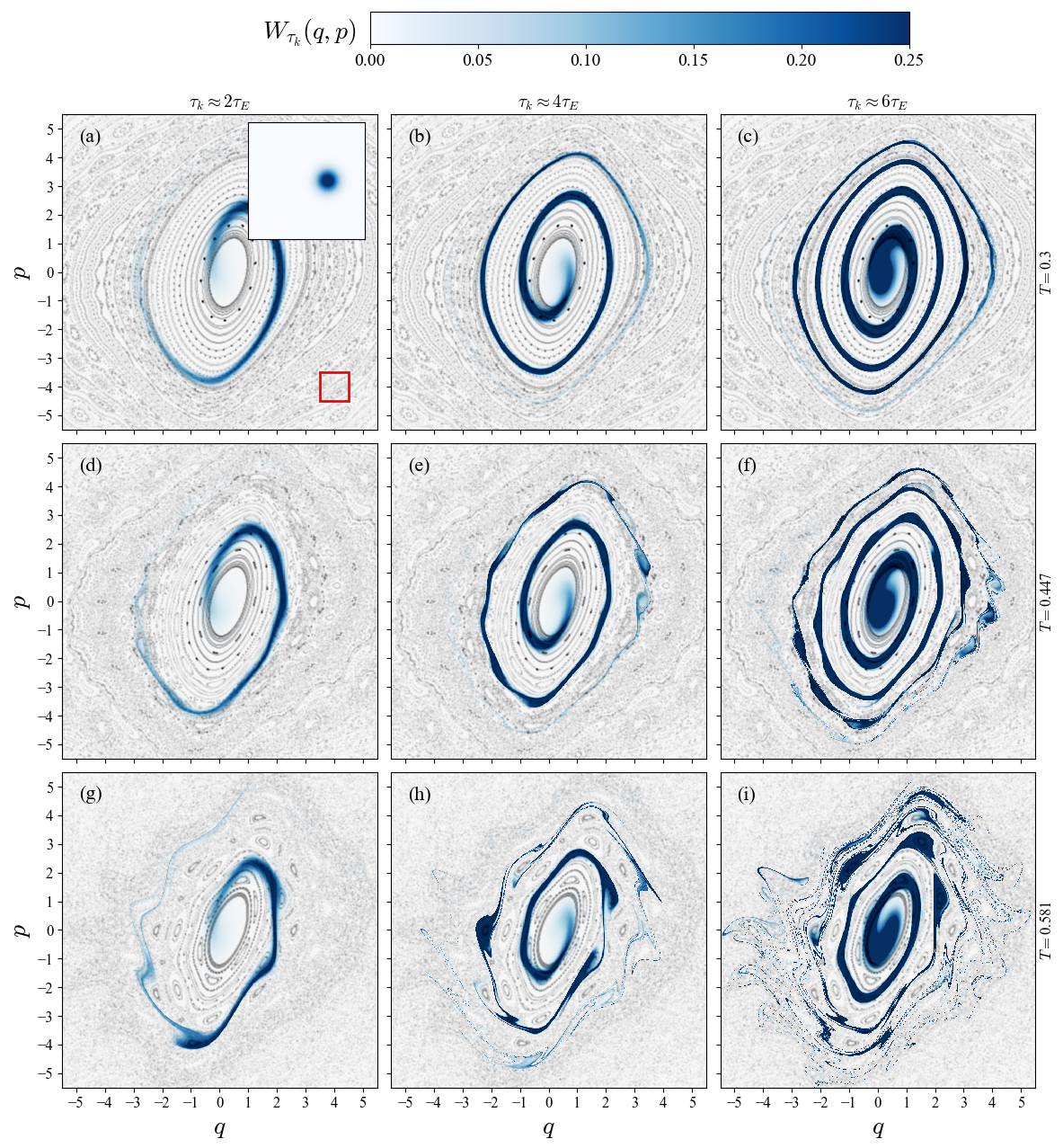}
	\caption{Classical propagation of the Wigner function shown in Fig.~\ref{fig:alpha}\textbf{(a)} by \eqref{eq:erfcosmap2} for $T = 0.3$, $T=0.447$ and $T=0.581$ and discretized times near $2\tau_E$, $4\tau_E$ and $6\tau_E$. Each $T$ value corresponds to a row, whereas each Ehrenfest time multiple corresponds to a column. The initial Wigner function is reproduced in the inset of panel \textbf{(a)}, together with a red square of area $\hbar=1$.  We also plot the classical map as a gray background in order show the orbits along which the initial Wigner function is propagated. Notice that the regular behavior for $T=0.3$ is gradually lost as $T$ is increased, the extreme case being the panels for $T=0.581$, where most of the orbits have become chaotic. The regimes for $T=0.447$ and $T=0.581$ have been considered problematic due to the filament developing sub-Planckian structure and being captured by stability islets \cite{Maitra2000}.}
	\label{fig:classical}
\end{figure*}

In Fig.~\ref{fig:classical} we display the classical evolution of the full Wigner function shown in Fig.~\ref{fig:alpha}(\textbf{a}) for the same kicking strengths as in Fig.~\ref{fig:alpha}(\textbf{b}), 1(\textbf{c}), and 1(\textbf{d}) (the time-scale used is described in Sec.~V.C). As expected, since the velocity varies with the initial point of an orbit, the Wigner function is deformed into a narrowing filament \cite{Lando2019}, which is scattered by fixed points along its way. The first row in Fig.~\ref{fig:classical} exemplifies the case of an initial Wigner function completely surrounded by regular trajectories, for which we expect the semiclassical approximation to be quite accurate. In the second row, with regular, bifurcated and weakly chaotic trajectories, we can clearly see some portions of the filament are captured by stability islands along the evolution. The third row deals mostly with chaotic orbits and we can see that the majority of phase space has been dominated be either chaos or sub-Planckian stability islands. For this latter case, the initial coherent state narrows into a thin filament that folds multiple times upon itself and runs over dozens of islands along its evolution, forming ``whorls" and ``tendrils" \cite{Berry1979-2}.

\subsection{The quantum coserf map and its semiclassical approximation}

\begin{figure*}
	\includegraphics[width=\linewidth]{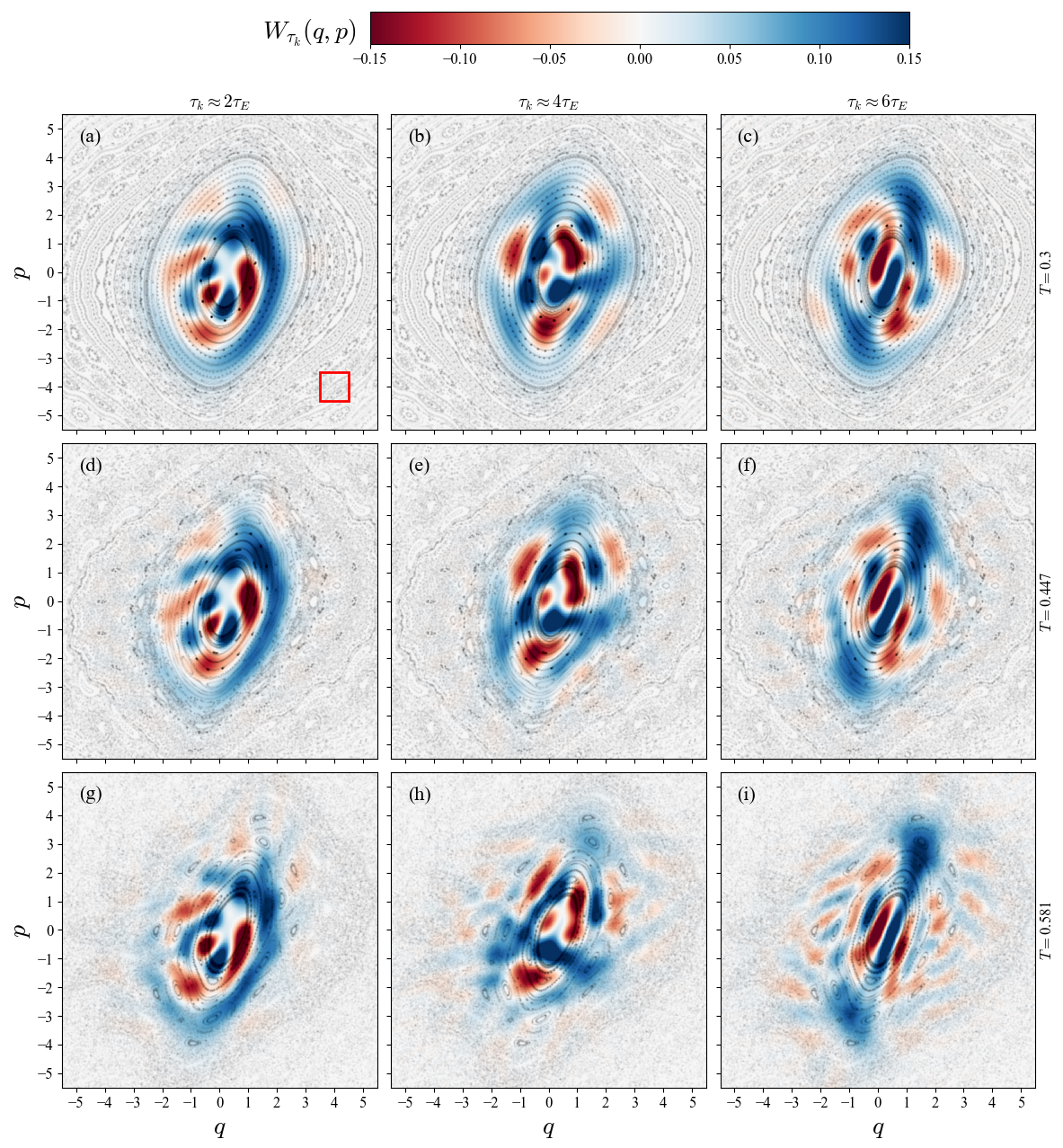}
	\caption{The quantum equivalent of Fig.~\ref{fig:classical} for the same values $T = 0.3$, $T=0.447$ and $T=0.581$ and discretized times $2\tau_E$, $4\tau_E$ and $6\tau_E$. The initial Wigner function is the same as in the inset of Fig~\ref{fig:classical}\textbf{(a)}. The red square in \textbf{(a)} has area equal to $\hbar = 1$, and the classical map is still plotted as a gray background. By comparing \textbf{(a)}, \textbf{(d)} and \textbf{(g)} with Fig.~\ref{fig:classical}\textbf{(a)}, \ref{fig:classical}\textbf{(d)} and \ref{fig:classical}\textbf{(g)} we notice that the quantum Wigner function has a discernible classical backbone for $\tau_k = 2 \tau_E$, around which quantum interference patterns are formed. Surprisingly, for $\tau_k = 6 \tau_E$ we have a quantum Wigner function that closely resembles a Sch\"odinger cat state, while its classical counterpart has simply narrowed and revolved multiple times around the origin. The Wigner functions for the other $T$ values look like deformations of the one for $T=0.3$, vaguely reflecting the changes suffered by the classical map's orbits as $T$ increases.}
	\label{fig:quantum}
\end{figure*}

For the quantum coserf map, we substitute the potential \eqref{eq:erfcosv} into \eqref{eq:qpos} and employ it to iterate an initial coherent state using \eqref{eq:qk}. The corresponding Wigner functions are calculated according to Sec.~III and are displayed in Fig.~\ref{fig:quantum} for the same kicking strengths introduced in Fig.~\ref{fig:alpha}. This figure should be seen as the quantum equivalent of Fig.~\ref{fig:classical} and uses the same time scale, to be introduced in the following section. The reader should observe the contrast between the classical and quantum scenarios by comparing Figs.~\ref{fig:classical} and \ref{fig:quantum}.

\vfill
\vfill

\vfill
\begin{figure*}
	\includegraphics[width=\linewidth]{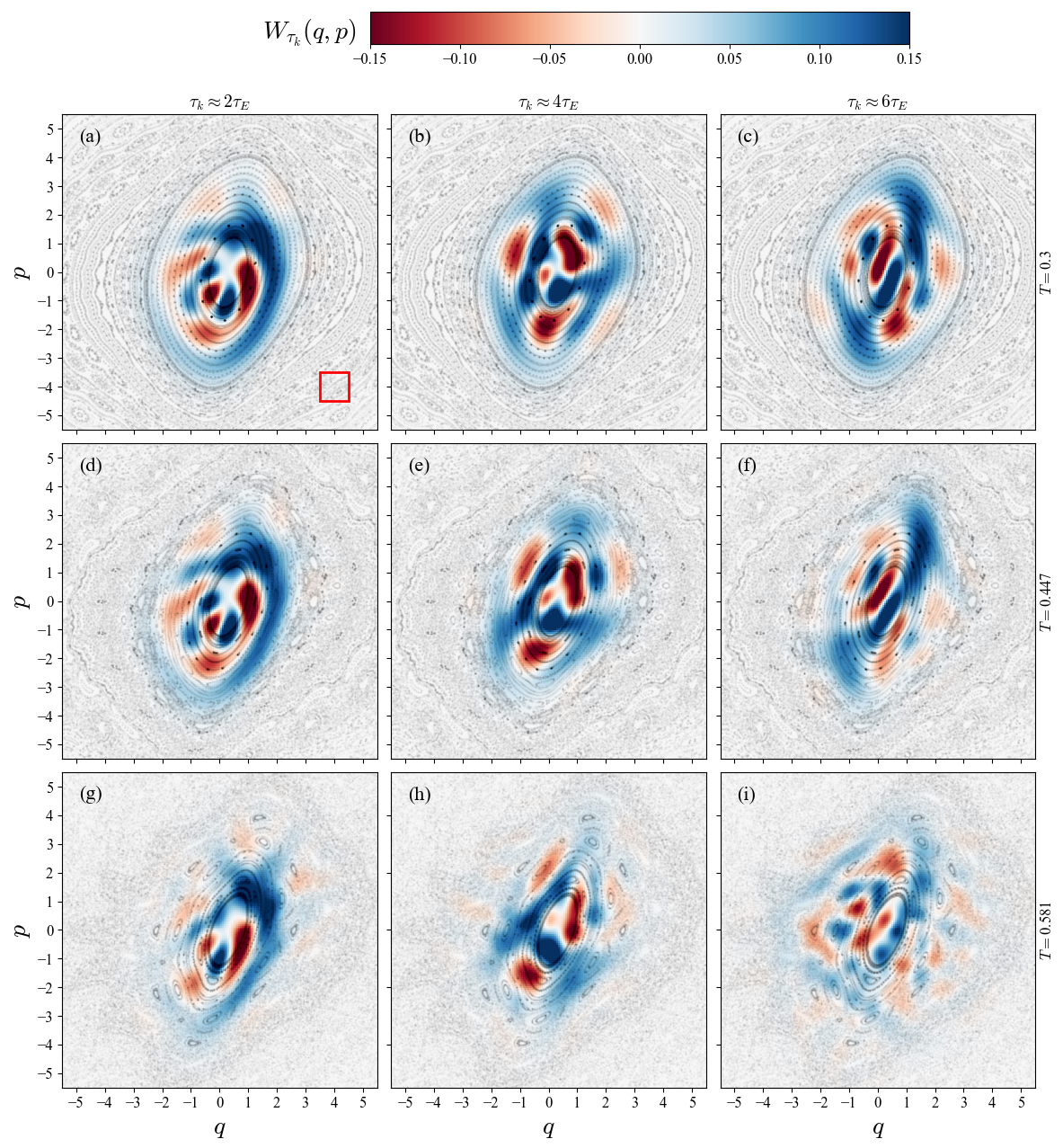}
	\caption{The semiclassical equivalent of Fig.~\ref{fig:classical} and Fig.~\ref{fig:quantum} for the same values $T = 0.3$, $T=0.447$ and $T=0.581$ and discretized times $2\tau_E$, $4\tau_E$ and $6\tau_E$. The initial Wigner function is the same as in the inset of Fig~\ref{fig:classical}\textbf{(a)}. The red square in \textbf{(a)} has area equal to $\hbar = 1$. For the completely regular case of $T=0.3$, the semiclassical approximations and the exact quantum Wigner functions are almost indiscernible (see also Fig.~\ref{fig:probs}). The stability islets present for $T=0.447$ are not enough to spoil the semiclassical approximation's accuracy even for the very long time $\tau_k = 6 \tau_E$, for which we can see that the filament has been captured by several islands. For the extreme $T=0.581$, the approximation finally breaks down for $\tau_k = 6 \tau_E$.}
	\label{fig:semiclassical}
\end{figure*}
\vfill

\begin{figure*}
	\includegraphics[width=\linewidth]{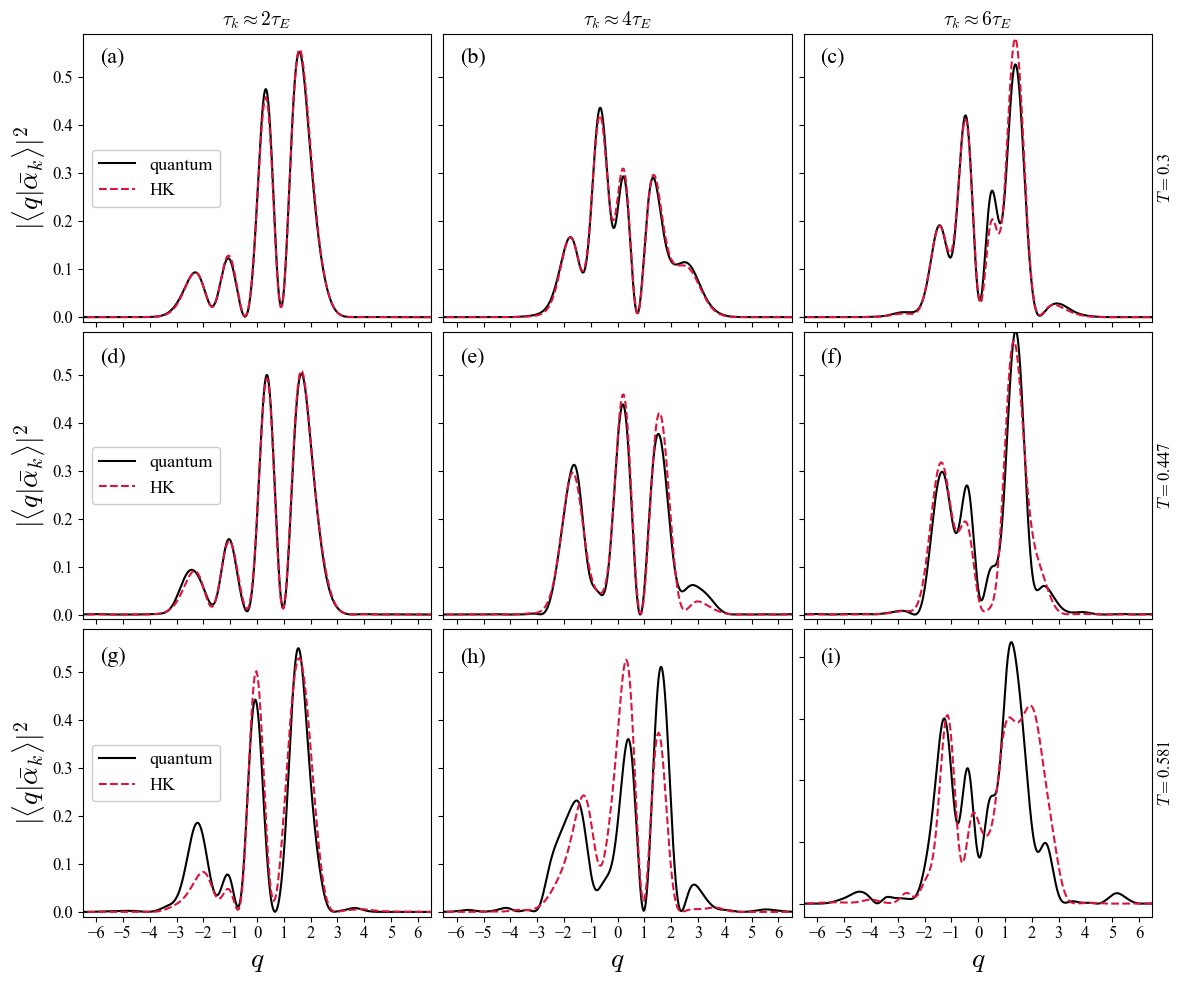}
	\caption{Position probability marginals obtained from the exact evolutions in Fig.~\ref{fig:quantum} and the renormalized semiclassical Wigner functions of Fig.~\ref{fig:semiclassical}.}
	\label{fig:probs}
\end{figure*}

\begin{figure*}
	\includegraphics[width=\linewidth]{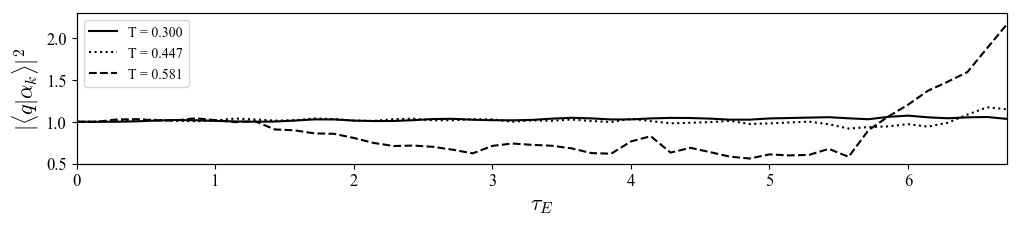}
	\caption{Normalizations for different kicking strengths, $T$, as functions of the Ehrenfest time $\tau_E = T k_E$. Although the kicks are discrete variables, we connect the points using lines for ease of visualization. For the $T = 0.3$ (solid line), we see in Fig.~\ref{fig:classical} that the initial wavepacket is still confined in a regular phase-space region and normalization is not lost. For $T=0.447$ (dotted line) we start to see deviations from unity around $\tau_k = 4 \tau_E$. The large number of chaotic orbits for $T=0.581$ (dashed line) causes an evident normalization loss for times just after $\tau_E$.}
	\label{fig:norm}
\end{figure*}

The semiclassical approximation to the coserf map is obtained according to Sec.~IV. As is well known, initial and final value representations such as Herman and Kluk's are bound to diverge when chaotic trajectories are present, even for systems defined on compact phase spaces such as the torus \cite{Liberto2016}. This may cause the wavefunction obtained via HK to lose normalization, a condition that is substantially worsened as time increases. This normalization loss was already evinced in the past (\emph{e.g.} \cite{Maitra2000}) and is not at all surprising, but it does not necessarily lead to the conclusion that the semiclassical approximation is no longer providing useful results. Instead of dismissing the method when the output starts to lose normalization, we define a renormalized wavefunction 
\begin{align}
\langle q \vert \bar{\alpha}_k \rangle = \dfrac{\langle q \vert \alpha_k \rangle }{\int dq \, \vert \langle q \vert \alpha_k \rangle \vert^2  } \, ,
\end{align}
such that 
\begin{align}
	\int dq \, \vert \langle q \vert \bar{\alpha}_k \rangle \vert^2  = 1 \, .
\end{align}
In Fig.~\ref{fig:semiclassical} we present the semiclassical equivalent of Figs.~\ref{fig:classical} and \ref{fig:quantum} as obtained from such a renormalized semiclassical wavefunction. The corresponding position probability marginals are shown in Fig.~\ref{fig:probs}. We also present the normalizations of $\langle q \vert \alpha_k \rangle$ in Fig.~\ref{fig:norm}.

\subsection{Ehrenfest time-scale and autocorrelation functions}

\begin{figure}
	\includegraphics[width=\linewidth]{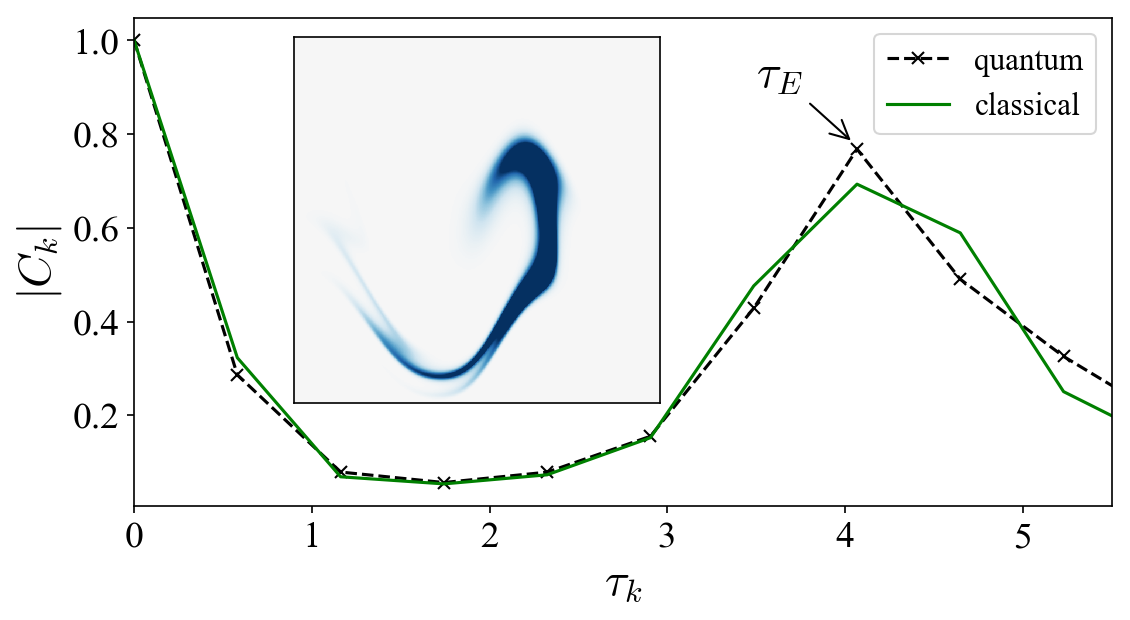}
	\caption{The meaning of the Ehrenfest time can be assessed through both autocorrelation function and phase-space geometry. Here we plot the quantum and classical autocorrelation functions for a coherent state initially centered at ($q=2,p=0)$ as a function of the discretized time $\tau_k = T k$, where $T=0.581$. The approximate moment at which these start to disagree is the Ehrenfest time $\tau_E$. In the inset  we display the classical Wigner function at $\tau_k = \tau_E$, showing it has performed roughly one full revolution around the origin.}
	\label{fig:autocorr}
\end{figure}

As classical trajectories intertwine, it is expected that their reproduction of quantum effects starts to become troublesome. In fact, the hypothesis that semiclassical methods should fail when tangles smaller than an $\hbar$-sized cell start to develop was consensual among most physicists until a series of papers in the 1990s showed that semiclassical propagation was accurate for much longer than this earlier estimate \cite{Heller1991, Tomsovic1991, Heller1992, Scharf1996}, although the bounds for the duration of this accuracy remain unclear. One of the earlier limits, the Ehfenfest time, is usually defined using Lyapunov exponents and characteristic actions for both integrable and chaotic systems \cite{Raul2012}. Instead, using the absolute value of the autocorrelation function 
\begin{align}
\vert C_k \vert = \vert \langle \alpha_0 \vert \alpha_k \rangle \vert \,, \quad C_0 = 1\, , \quad k \in \mathbb{N} \, ,  \label{eq:autoquan}
\end{align} 
the Ehrenfest time $\tau_E$ can be defined as the instant at which the classical and the quantum expressions from $\vert C_k \vert$ no longer agree, the latter developing oscillations due to quantum interferences. Since the coserf map has an elliptic fixed point at the origin, $\tau_E$ will also correspond to the time at which the classically propagated Wigner function performs a complete turn around the origin \cite{Raul2012, Lando2019}. These ideas are depicted visually in Fig.~\ref{fig:autocorr}.

\begin{figure*}
	\includegraphics[width=\linewidth]{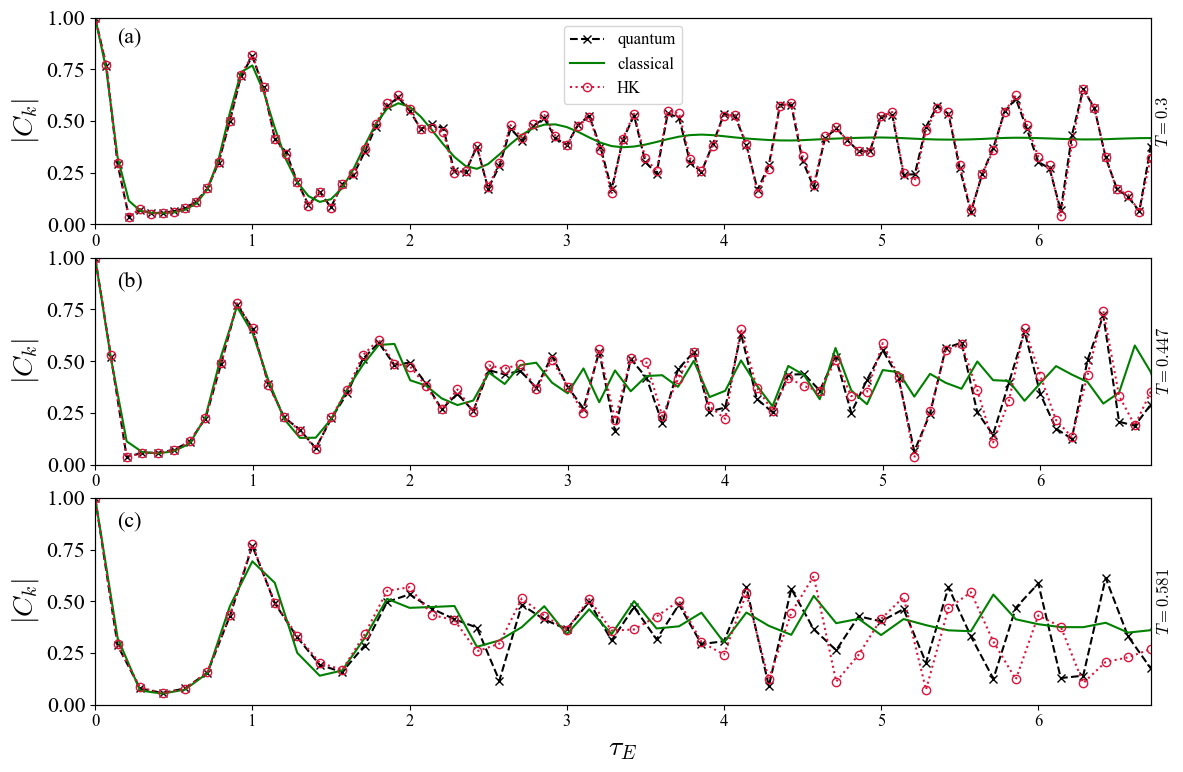}
	\caption{Classical, quantum and semiclassical autocorrelation functions obtained respectively from \eqref{eq:autoclass} and \eqref{eq:autoquan} using both the exact evolution \eqref{eq:qkick2} and its HK approximation, obtained from substituting \eqref{eq:HK} in \eqref{eq:qk}. As in Fig.~\ref{fig:norm}, we use the Ehrenfest time scale and the initial coherent state is centered at $(q=2,p=0)$. The autocorrelations are discrete entities, but we connect them using lines in order to facilitate interpreting the plots. Notice how the semiclassical autocorrelations only start to fail near the time values for which the normalizations in Fig.~\ref{fig:norm} also start to oscillate.}
	\label{fig:autocorr2}
\end{figure*}

Since $\vert \alpha_k \rangle$ is a function of the kicking strength $T$, so is the Ehrenfest time: As $T$ increases, the ``Ehrenfest kick" $k_E$ at which the classical and quantum autocorrelations separate becomes increasingly shorter, but $\tau_E$ itself remains constant and equal to $T k_E$, \emph{i.e}, $\tau_E/k_E = T$. This is just another way to state that as chaotic orbits start to dominate phase space, the time for which the semiclassical approximation should remain valid starts to decrease according to this \emph{a priori} criterion. In Fig.~\ref{fig:autocorr2} we display the quantum, classical and renormalized semiclassical autocorrelation functions for the kicking regimes of Figs.~\ref{fig:classical}, \ref{fig:quantum} and \ref{fig:semiclassical}.

\section{Discussion}

It is well known that quantum mechanics ``washes over" classically convoluted regions, especially when dealing with the propagation of coherent states \cite{Gutzwiller1990,Tomsovic1991,Berry1979}. Schr\"odinger's equation, being linear, is somewhat blind to the intricacies attributed to the chaotic nature indigenous to classical mechanics. We have already observed in Figs.~\ref{fig:semiclassical} and \ref{fig:probs} how some of the filaments in Fig.~\ref{fig:classical}, with unmistakable influence from chaotic orbits and stability islands, were able to reproduce almost flawlessly their quantum equivalents in Fig.~\ref{fig:quantum}. It is no surprise that thin, regular filaments are able to semiclassically recreate intricate quantum interference patterns, a phenomenon that was already evinced as the consequence of constructive and destructive interferences between phases associated to classical trajectories \cite{Lando2019}. Indeed,  by comparing the first lines of Figs.~\ref{fig:classical} and \ref{fig:semiclassical} we see that this phase matching is so perfect that the semiclassical approximation is able to reproduce the quantum result using classical trajectories that lie in a region where the final Wigner function is, in fact, zero. 

It is unexpected, however, that such good agreements between quantum and semiclassical mechanics should be attained when the classical filament has been captured by so many stability islands, such as in Figs.~\ref{fig:classical}\textbf{(e)} and \ref{fig:classical}\textbf{(f)}. Nevertheless, if we take a closer look at Figs.~\ref{fig:classical}\textbf{(e)} and \ref{fig:probs}\textbf{(e)} we can associate the only region where the semiclassical marginal is slightly off, namely at the interval $q \in (2,4)$, to the presence of filamentary classical structure. This scenario is also visible for Fig.~\ref{fig:classical}\textbf{(g)} in $q \in (-3,-1)$, and for Fig.~\ref{fig:classical}\textbf{(h)} in $q \in (2,3)$. The latter presents the interesting phenomenon that the chain of islands around $q=4$ is completely canceled in the marginal, which is quite accurate at this value for $q$. 

Before associating failures to islands, one should again take a look, for instance, at Figs.~\ref{fig:classical}\textbf{(f)}: Notice how the region $q \in (-3,-1)$ is full of islands and captured filamentary pieces, but the corresponding marginal in Fig.~\ref{fig:probs}\textbf{(f)} is extremely accurate in this interval. Somehow, the classical structure that forms for $q < -2$ is also canceled by the semiclassical approximation. The autocorrelation functions of Fig.~\ref{fig:autocorr2}, with the exception of the strongly chaotic regime, are also remarkably accurate despite the island chains. It is therefore possible that certain types of sub-Planckian islands might affect semiclassical accuracy, while others do not impact it. A distinction might also exist between the cases where we expect the final classical contribution to be canceled or to be different from zero. Of course, the classical structure's thickness certainly plays a major role, since the contributions due to thin filaments are expected to vanish, as we can see in Figs.~\ref{fig:classical}\textbf{(h)} and \ref{fig:semiclassical}\textbf{(h)} at $q=3$. 

Leaving the quantitative terrain of probability marginals and autocorrelations, the main achievement of this work was to present, for the first time, a geometrical means to visualize quantum mechanics as emerging from classical trajectories for a highly non-trivial system. Despite the inaccuracies visible in Figs.~\ref{fig:autocorr2} and \ref{fig:probs}, the reader should dedicate some attention to Fig.~\ref{fig:semiclassical} and its unarguable success: With the exception of the \textbf{(i)} panel, all of the fundamental quantum structure was semiclassically captured and, for many panels, flawlessly reproduced. Even if some fine details were not assessed, it is hard to imagine that, \emph{e.g.} an experimental realization of some system that obtains Fig.~\ref{fig:semiclassical}\textbf{(h)} instead of Fig.~\ref{fig:quantum}\textbf{(h)} is not successful. Even when the approximation fails, as in the \textbf{(i)} panels of Figs.~\ref{fig:quantum} and \ref{fig:semiclassical}, the HK approximation was able to more or less pinpoint the correct peaks, especially if one considers its input in \ref{fig:classical}\textbf{(i)}. (At this point it is fundamental to state that extracting meaningful results from strongly chaotic regimes is not a straightforward process and is discussed in Appendix B, where a better alternative to Fig.~\ref{fig:semiclassical}\textbf{(i)} is presented.)

We remind the reader that this work is based on a very long time scale, semiclassically speaking, and that for $\tau_k < 2 \tau_E$ our semiclassical results are almost identical to the exact quantum case for all kicking strengths tested.

\section{Conclusion}

We have carried out a comprehensive semiclassical analysis of the evolution of an initial coherent state under the action of a softly chaotic map on the plane. The semiclassical results were obtained using the Herman-Kluk propagator. The time regimes encompass short, intermediate and long times alike. Comparisons between classical, quantum and semiclassical results relied on an extensive apparatus: Probability marginals, Wigner functions and autocorrelations. Geometrical analogies were vividly drawn from a phase-space perspective, while the quantitative analysis proved the semiclassical approximation to be remarkably accurate, providing useful information even when the underlying classical dynamics has become heavily populated by sub-Planckian structure. The approximation's eventual failure was also attributed to the presence of strongly chaotic orbits, the thinning of classical filaments and, sometimes but not always, stability islands. In the particular case of regular dynamics the semiclassical approximation did not seem to ever fail -- even for times much longer than a dozen Ehrenfest times.

\section*{Acknowledgements}

Partial financial support from CNPq
and the National Institute for Science and Technology:
Quantum Information is gratefully acknowledged.

\section*{Appendix A: Computational and numerical details}

For the exact quantum wavefunction we use a position grid with $N=2^{13}$ equally spaced points, with spacing $\Delta q = \sqrt{\pi/N}$. The grid-size $N$ is taken as a power of 2 because this maximizes the number of intrinsic symmetries in the Discrete Fourier Transform (DFT) algorithm, which we use to Wigner-transform the wavefunction \footnote{See, for instance, the documentation of \texttt{fft} at docs.scipy.org/doc/numpy/reference/routines.fft.html.}. This choice of $\Delta q$ is also required in order to employ the DFT, and although unnecessarily large grids to attain convergence are used, this is compensated by the algorithm's efficiency.

Semiclassical calculations were carried out on a $501 \times 501$ equally spaced phase space grid, with position and momentum running from $-4\pi$ to $4\pi$. This implies $\Delta q \, \Delta p = (\Delta q)^2 \approx 2.5 \times 10^{-3}$ and approximately $5 \times 10^3$ trajectories used for integration. The autocorrelation function and normalization for $T=0.581$ were also evaluated on these large grids, but for $T=0.3$ and $T=0.447$ we halved grid-size: This was already enough to restrict deviations to $\mathcal{O}(10^{-3})$. 

The numerical analysis in this paper was all implemented in Python 3.0 and interpreted on a notebook running the Linux Mint 18.3 OS. The relevant specifications of this machine are its dual core, 2x2.7 GHz i7-7500U CPU and its 8Gb of RAM.  

\section*{Appendix B: Sensitivity to initial grids}
${}$
\begin{figure}
	\includegraphics[width=\linewidth]{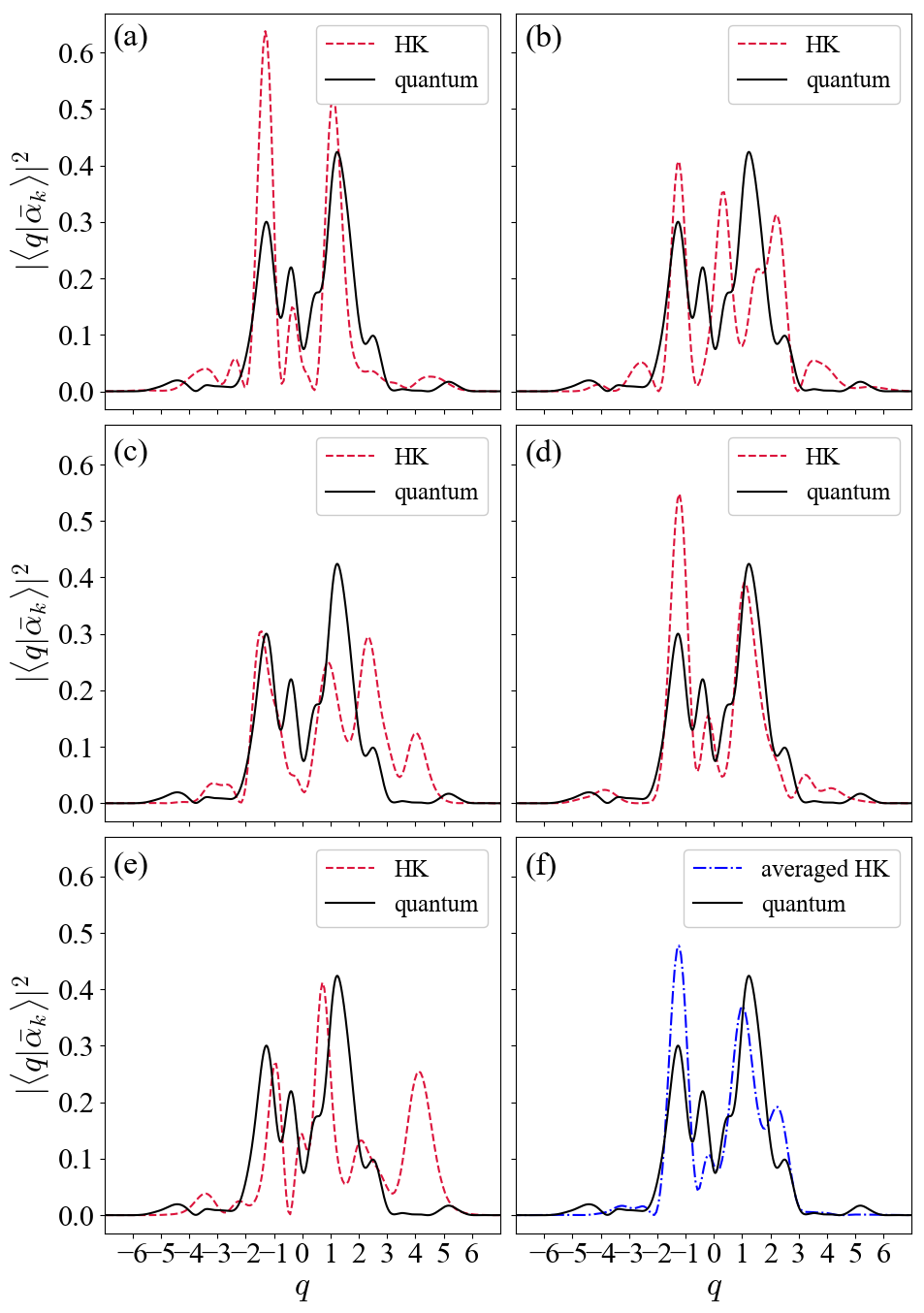}
	\caption{Quantum and semiclassical renormalized position marginals for an initial coherent state centered at $(q_0=2, p_0=0)$. We adopt the same parameters as in Fig.~\ref{fig:probs}\textbf{(i)}, \emph{i.e} $T=0.581$ and $\tau_k = 6 \tau_E$. The grid spacings are: \textbf{(a)} $(\Delta q)^2 \approx 0.0726$; \textbf{(b)} $(\Delta q)^2 \approx 0.0722$; \textbf{(c)} $(\Delta q)^2 \approx 0.0718$; \textbf{(d)} $(\Delta q)^2 \approx 0.0714$; \textbf{(e)} $(\Delta q)^2 \approx 0.0701$. The simple average of all these marginals is displayed in \textbf{(f)}.}
	\label{fig:grids}
\end{figure}

An often ignored phenomenon when dealing with chaotic systems is that the sensitivity to initial conditions is reflected when choosing not only the state to be propagated, but also the initial numerical grid: Slightly changing the grid spacing of an initial position and momentum mesh can significantly impact the final grid after the onset of strong chaos. This is rather obvious, since changing grid sizes implies changing initial conditions. We must, therefore, be mindful about the possibility of choosing grids (and, therefore, trajectories) that might improve or worsen our semiclassical approximations. We exemplify this in Fig.~\ref{fig:grids} by reproducing Fig.~\ref{fig:probs}\textbf{(i)} using several initial grids, all of which have almost the same spacing, and observing the impact of grid size in the semiclassical probability marginal. Here, the effect of an $\mathcal{O}(10^{-4})$ variation in $(\Delta q)^2$ creates a completely different semiclassical marginal due to the sensitivity of the underlying chaotic dynamics. A more stable semiclassical result is also presented in Fig.~\ref{fig:grids}\textbf{(f)} by taking the simple average over the earlier panels. Such improved averaged output is much more reliable and is the subject of current investigation.

The stark deviations of Fig.~\ref{fig:grids} are not seen for any other values of kicking strength presented in this paper and, in fact, no change at all is observed for $T=0.3$ and $T=0.447$. This is expected, since these kicking regimes are not composed of as many chaotic orbits as $T=0.581$. \\
{}

\bibliographystyle{apsrev4-1}
\bibliography{maps}

\end{document}